\documentclass[a4paper, 10pt, conference]{IEEEtran} 

\DeclareMathAlphabet\mathbfcal{OMS}{cmsy}{b}{n} 
\usepackage{cite}
\usepackage[pdftex]{graphicx}
\usepackage{dsfont}
\usepackage{comment} 
\usepackage[usenames,dvipsnames]{color}
\usepackage{acronym}
\usepackage{multirow}
\usepackage[caption=false]{subfig}
\usepackage{nth}
\usepackage[geometry]{ifsym}

\usepackage{amsfonts}
\usepackage{amsmath,amssymb}
\usepackage{wasysym} 
\usepackage{mdwmath}
\usepackage{mdwtab}
\usepackage{balance}
\usepackage[hyphens]{url}
\usepackage{placeins}
\usepackage{xcolor, soul}
\sethlcolor{SkyBlue}
\usepackage{float}
\usepackage{amsmath}

\DeclareMathOperator*{\argmin}{arg\,min}
\ifCLASSINFOpdf
\else
\fi 
\begin{document}

\def\user{i}  
\def\userOth{j}
\def\player{\mathcal{I}}
\def\numPlayer{N}
\def\strategy{S}
\def\strategyI{s_{\user}}
\def\strategyJ{s_{\userOth}}
\def\payoff{U}
\def\reelnum{\mathds{R}}
\def\game{\mathcal{G}}
\def\bestresp{\mathcal{B}}
\def\noise{\omega(t)}
\def\totalpower{\mathcal{P}}
\def\mypower{\psi}
\def\pu{\mathcal{M}}
\def\su{\mathcal{F}}
\def\rx{\mathcal{P}}
\def\supower{\psi}
\def\antennagain{\rho}
\def\infobit{{L}}
\def\datarate{{R}}
\def\infot{{M}}
\def\sinr{\gamma_{\user}}
\def\stateInd{H}
\def\subcspace{\Delta f}
\def\landa{\h}
\def\in{∈}

\def\offRate{\varphi}
\def\rollOff{\beta}
\def\potentialG{V}
\def\utility{\payoff(\offRate,\rollOff)}
\def\equSinr{\frac{p_{\user}}{p_j+\noise_0}}
\def\equDer{\frac{p_{\user}}{(p_j+\noise_0)^2}}

\newtheorem{definition}{Definition}
\def\figurePath{pics/}
\graphicspath{ {pics/} }
\def\figWidth{3.45}
\makeatletter
\newcommand*{\rom}[1]{\expandafter\@slowromancap\romannumeral #1@}
\makeatother
\newcommand{\vect}[1]{\boldsymbol{#1}}
\providecommand{\keywords}[1]
{
	\small	
	\textbf{\textit{Keywords:}} #1
}

\title{Predicting the Path Loss of Wireless Channel Models Using Machine Learning Techniques in MmWave Urban Communications
}

\author{\IEEEauthorblockN{Saud Aldossari, Kwang-Cheng Chen}\\

\IEEEauthorblockA{Department of Electrical Engineering University of South Florida Tampa, Florida 33620 \\ Email:  Saldossari@mail.usf.edu, kwangcheng@usf.edu }}

\maketitle
\begin{abstract}
The classic wireless communication channel modeling is performed using Deterministic and Stochastic channel methodologies. Machine learning (ML) emerges to revolutionize system design for 5G and beyond. ML techniques such as supervise leaning methods will be used to predict the wireless channel path loss of a variate of environments base on a certain dataset. The propagation signal of communication systems fundamentals is focusing on channel modeling particularly for new frequency bands such as MmWave. Machine learning can facilitate rapid channel modeling for 5G and beyond wireless communication systems due to the availability of partially relevant channel measurement data and model. When irregularity of the wireless channels leads to a complex methodology to achieve accurate models, appropriate machine learning methodology explores to reduce the complexity and increase the accuracy. In this paper, we demonstrate alternative procedures beyond traditional channel modeling to enhance the path loss models using machine learning techniques, to alleviate the dilemma of channel complexity and time consuming process that the measurements take. This demonstrated regression uses the measurement data of a certain scenario to successfully assist the prediction of path loss model of a different operating environment.
	\end{abstract}


\acrodef{icic}   [ICIC]   {inter-cell interference coordination}
\acrodef{ne}     [NE]     {Nash equilibrium}
\acrodef{ofdma}  [OFDMA]  {orthogonal frequency-division multiple access}
\acrodef{cdma}   [CDMA]   {code-division multiple access}
\acrodef{ofdm}   [OFDM]   {Orthogonal frequency-division multiplexing}
\acrodef{rbs}    [RB]     {resource block(s)}
\acrodef{lte}    [LTE]    {long term evolution}
\acrodef{wimax}  [WiMAX]  {worldwide interoperability for microwave access}
\acrodef{sinr}   [SINR]   {signal to interference plus noise ratio}
\acrodef{ra}     [RA]     {resource allocation}
\acrodef{bs}     [BS]     {base station}
\acrodef{ue}     [UE]     {user equipment}
\acrodef{gt}     [GT]     {Game theory}
\acrodef{hn}     [HetNets]{heterogeneous networks}
\acrodef{sg}     [SG]     {supermodular game}
\acrodef{cdf}    [CDF]    {cumulative distribution function}
\acrodef{ue}     [UE]     {user equipment}
\acrodef{fbs}    [FBS]    {femto BS}
\acrodef{rrh}    [RRH]    {remote radio heads}
\acrodef{henb}   [HeNB]   {enhanced Home Node B}
\acrodef{fue}    [FUE]    {femto user equipment}
\acrodef{mue}    [MUE]    {macro user equipment}
\acrodef{su}     [SU]     {secondary user}
\acrodef{pu}     [PU]     {primary user}
\acrodef{pbs}    [PBS]    {primary base station}
\acrodef{sbs}    [SBS]    {secondary base station}
\acrodef{rss}    [RSS]    {received signal strength}
\acrodef{tdd}    [TDD]    {time division duplexing}
\acrodef{awgn}   [AWGN]   {additive white Gaussian noise}
\acrodef{pot}    [POT]    {Partially overlapping tones}
\acrodef{pofmt}  [POFMT]  {partially overlapping filtered multi tones}
\acrodef{fmt}    [FMT]    {Filtered multi-tones}
\acrodef{pnw}    [P'nW]   {Play\&Wait}
\acrodef{cop}    [CoP]    {continuous play}
\acrodef{rrc}    [RRC]    {root raised cosine}

\renewcommand{\figurename}{Fig.} 
\renewcommand{\tablename}{Table }
\def\equname{Equ.}

\IEEEpeerreviewmaketitle


\section{Introduction}
\label{intero.}
\IEEEPARstart{S}{tate-of-the-art} channel modeling is the process of predictively incorporating wireless channel parameters into a channel model using minimal number of measurements. Radio propagation models can be traditionally obtained via Deterministic and Stochastic Channel Models and applying a regular statistical method to build a model. These traditional methods are becoming more complex and time consuming by using the new measurements while employing new technologies/frequency-bands and the increase of data traffic \cite{Ericsson2011}\cite{Sultan}. In this paper, we are introducing a new procedure to overcome this dilemma of channel modeling by machine learning (ML) that emerges to revolutionize systems design for 5G and beyond. ML techniques, particularly supervise leaning methods, will be used to predict the wireless channel path loss, a key component of channel modeling. As mmWave frequency bands are widely introduced to 5G and thus require tremendous of new measurements, the irregularity of the wireless channels leads to a complex methodology in order to to achieve accurate models. Machine learning algorithms therefore aim to reduce such complexity and increase the accuracy while reduces the number of measurements. From the computational aspect, channel modeling can be considered as a sort of data mining and machine learning techniques considered as a valid solution to predict models instead of empirical and deterministic methods \cite{Piacentini2011}\cite{cwang2018}. \cite{Timothy2017} applied machine learning techniques to predict the carrier frequency offset (CFO) and presented better result toward applying machine learning to the wireless channel modeling.
The prediction of channel model emerges as a critical mission in modern AI assisted communication systems. Furthermore, ML assists the extraction of useful information from the vast amount of channel measurement data in the wireless communication system \cite{Samuel1959}. Investigation of machine learning methods is suitable and capable to derive a channel modeling in a better shape than the traditional ways as other researchers have accomplished and recommended \cite{Zhao2013}. 
\begin{figure}[!h]
	\centering
	\setlength{\abovecaptionskip}{0.cm}
	\setlength{\belowcaptionskip}{0.00cm}
	\includegraphics[width=0.50\textwidth]{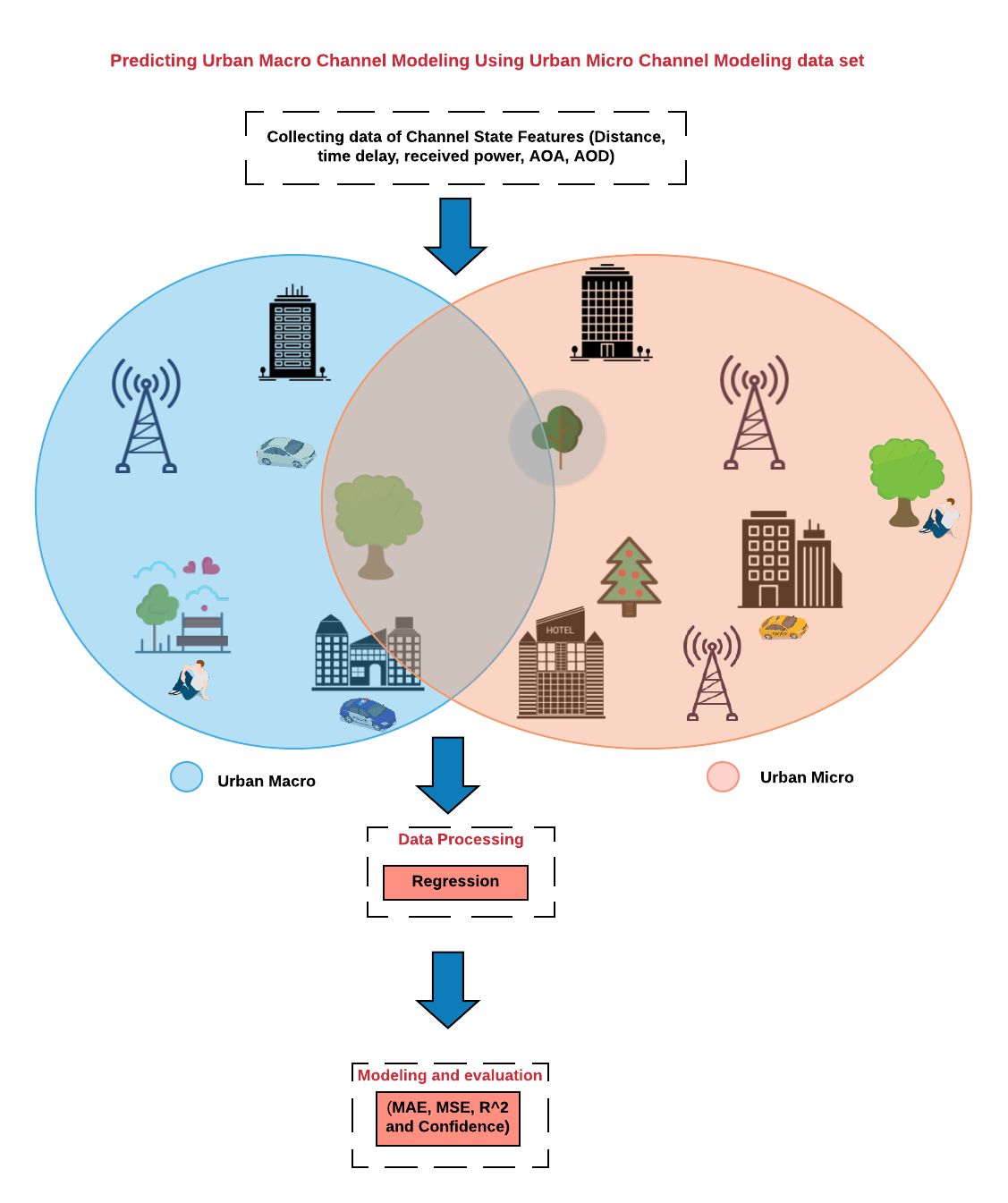}
	\renewcommand\figurename{Fig}
	\caption{The propose idea to reduce the number of measurement of wireless channel modeling.}
	\label{Fig1}
\end{figure}
Figure 1 illustrates the proposed idea of this journey, which can be seen as instead of applying a specific measurement campaign to obtain a certain wireless channel model. ML can be used to predict a model of variant of environments to predict a wireless channel model base on a reliable data from a different environment. The traditional way is accomplished by conducting tremendous of measurement in a particular environment and obtaining its model using regular statistical techniques. The general mathematical model can be shown as:
\begin{equation}
y(t)=x(t)*h(\tau,t)=\int_{-\infty}^{\infty}h(\tau,t)x(t-\tau)d\tau
\end{equation}
Where y(t) is the received signal, u(t) is the transmitted signal, * is the convolution sign and $h(\tau,t)$ is the delay spread function with respect to delay and time \cite{channelmodelbookk}. The characteristics of the wideband channel such the power delay profile PDP, RMS delay spread, and others channel parameters are derived from the channel impulse response $h(\tau,t)$.  The received passband signal is shown below:
\begin{equation}
Y(d,t)=\sum_{i=1}^{L-1}\alpha_i s(t-\tau_i)+n(t)
\end{equation}
Large scale fading (LSF) usually due to the object that shadow the signal and explains the main characteristics of the channel such as path loss, shadow, angular spread, etc. Moreover, LSF cases will be examined the relationship between the path loss and the separated distance between the Tx and Rx in different environments such as suburban.
\begin{equation}
APDP(\tau)=\frac{1}{N}\sum_{i=1}^{L}|h(t-\tau_i)|^2
\end{equation}
In this manuscript, common solutions to model large scale path loss is introduced in \rom{2}. Followed by section \rom{3} demonstrate how machine learning will be involved in wireless channel modeling. Then, generating a dataset of this work explained in section \rom{4}. Then models validation and results in \rom{5}. Lastly, a conclusion is shown in section \rom{6}.

\section{Conventional Model of Large Scale Path Loss}
The common ways to predict the path loss in a channel model vary base on different characteristics, such as environments, types of antenna and frequency scales. Inferring the path loss of a different communication environment using existing path loss models or data from different environments have not been well investigated in the literature, which serves our target and novelty with the aid of ML. There are non-ML types of path loss models that are used to predict the signal loss of the propagated link via a wireless channel. These path loss models are briefly explained in the following subsections. 
\subsection{Close-in (CI) Model}
This model is usually for LOS and NLOS for all urban micro (UMi), UMa, and InH as well by using close-in reference distance base on Fariis' law. The general version of the close-in model is \cite{5GCM2016}:
\begin{align}
\small{PL^{CI}(f,d)[dB]=FSPL(f,1,m)+10nlog_{10}(\frac{d}{d_o})+X^{Cl}_\sigma,}
\end{align}
The parameter FSPL is just the free space model in dB, $n$ is the path loss exponent (PLE) to show how PL varies with multipath propagation distance and $d_o$ is the reference distance which set to 1 m since there is rarely shadowing in the first meter and simplifies the equation as well. 
The above PL models should be applicable to our usage in the work and has the form of a linear regression model where other path loss models such as the following models can be applied to multiple linear regression due to the many channel features.
\subsection{Close-in with frequency dependent exponent}
This model was proposed in the 3GPP meetings where it depends on how high is the frequency. Close-in with frequency dependent is an extension of the CI model with frequency dependent \cite{Maco2016}. The general model of the CIF model shown as \cite{RanganM2014}.
\begin{multline}
PL^{CIF}(f,d)=FSPL(f,1,m)+10n(1+b(\frac{f-f_o}{f_o}))\\log_{10}(\frac{d}{1m}) +X^{CIF}_\sigma 
\end{multline} 
Note: $b$ is a parameter that captures the slop or the dependence of the path loss of the weighted average of the reference frequencies $f_0$  and it would be positive if the both PL and f increase. 

\cite{ShusunM2016} compared the above path loss model for both $\mu$Wave and MmWave in different scenarios such as UMa and UMi. All of them showed a good prediction with large data where the CI model was the most suitable for outdoor cases due to the close-in free space reference. Whereas, CIF has a better performance for the indoor environments due to its small stander deviation values. The path loss exponent in CI/CIF models shows loss with distance for urban macro then urban micro and that seems applicable due the obstructions blocking the signal from than lower base stations while urban macro is commonly higher than the micro communication.
\subsection{floating-intercept (FI)}
FI model is also called alpha-beta (AB) path loss model. This PL model can be combined with log-normal shadowing as shown:
\begin{equation}
PL^{FI}(f,d)[dB]=\alpha +10\beta log_{10}(d)+X^{FL}_\sigma 
\end{equation}
The values of $\alpha$  and $\beta$ can be obtained using the least square fitting as a slope and floating intercept respectively. Also, the shadow fading is represented by $X^{FL}_\sigma$ fallowing a Gaussian random variable with zero mean and standard deviation $\sigma$.
\subsection{Alpha-Beta-Gamma (ABG) Model}
ABG model is the current 3GPP 3D model and values may change based on the base station location.  
\begin{equation}
PL^{ABG}(f,d)[dB]=10\alpha log_{10}+\beta +10\gamma log_{10}(\frac{f}{1GHz}) \\ +X^{ABG}_\sigma 
\end{equation}
ABG model is used to measure how PL get increased with distance and $\alpha$ is the slope of PL with log distance. $\beta$ is the optimized floating offset in dB, $\gamma$ is the PL variation dependence over a frequency in GHz and $X^{ABG}_\sigma$ is the fading (SF) in dB. Since there are three parameters, 3GPP mentioned that ABG PL model always has a lower shadow fading standard deviation than other PL models \cite{ShuM2016}.

\section{Modeling the Path Loss Using Machine Learning Techniques}
Estimating a path loss can be solved by machine learning techniques to overcome challenging issues such as complexity and time consuming due to the required tremendous measurements. The classic wireless communication channel modeling is performed using Deterministic and Stochastic channel methodologies. In the following, ML techniques such as supervise leaning methods will be used to predict the wireless channel path loss of a variate of environments base on a certain dataset, with application scenarios like mmWave bands and vehicle-to-vehicle communication channels. In this paper, we demonstrate applying machine learning to develop alternative procedures to enhance the path loss models using machine learning techniques. Furthermore, investigation of machine learning methods is suitable and capable to derive a channel modeling in a better shape than the traditional ways as other researchers have accomplished and recommended \cite{Zhao2013}.
This section will presents how to apply ML methods to estimate the channel parameters using different regression methods. Regression is a basic supervise learning technique \cite{LuD2018}. Regressions use least square error (LSE) to minimize the square of the error between the observed responses in the dataset and to predict the most accurate model \cite{RiihijarviF2018} \cite{ZhangYan2018}. Regression is one of the main methods that is used in machine learning where the regression models learn the mechanism based on a dataset from prior measurements or simulations. After the learning processes are completed, the model coefficients can be obtained. Authors of \cite{LuD2018} applied support vector and DNNs regressions then, made a compression to control the high-speed channel modeling errors. Where all of these techniques help reducing the wireless channel modeling complexity and need to be investigated more.
The machine learning techniques that will be used in this work are linear and multiple linear regression algorithms. Multiple linear regression technique takes the advantages of other channel modeling features to enhance the path loss prediction comparing to regular linear regression and that can be seen in the future result section. Furthermore, this journey will test how the wireless channel features affects the path loss prediction. The reason of using regression techniques instead of other machine learning methods is due the desire of prediction continuous values.
\begin{equation}
Y_i = f_i(\vect{X}) + \epsilon_i
\end{equation}
$Y$ is the dependent response which is in our case the path loss, X is the independent variable in form of $ \vect{X} = \left[x_1,x_2,\ldots,x_p \right]$ which is the channel state information (CSI) features such as distance, time delay, received power, azimuth AoD, elevation AoD, azimuth AoA, RMS Delay Spread, and frequency (GHz). In order to make a prediction of $Y$ on new data, we need to estimate $\hat{f}(X)$. Thus, the estimate coefficients have to be accurate as possible to enhance the accuracy.
Path loss models suppose to be applicable and have the form of regression algorithms such as linear regression for Floating-Intercept (FI) model. While other path loss models such as the Alpha-Beta-Gamma (ABG) model can be applied to multiple linear regression due to the other channel features in the previous section . By considering linear regression where distance is the only channel feature used to estimate the path loss model as show below and estimating the parameters can be seen in equation 14.
\begin{equation}
\hat{Y}=\hat{\beta_0}+\sum_{i=1}^{k} \hat{\beta_i} X_i
\end{equation}
\begin{equation}
E(\beta_0,\beta_i)=\sum_{i=1}^{p}[y_i-(\beta_0+\beta_iX_i)]^2
\end{equation}
By applying this approach to estimate the the coefficient parameters and then characterizing the theoretical loss $L$ is obtained. 
\begin{equation}
\argmin_jL=\frac{1}{N}\sum(\hat{Y}(k)-Y(k))^2
\end{equation}
\subsection{Multiple Linear Regression}
Multiple linear regression technique takes the advantages of other channel modeling features to enhance the path loss prediction comparing to regular linear regression. Furthermore, this journey will test how the wireless channel features affects the path loss prediction. 
Machine learning techniques would be adopted to estimate the channel modeling parameters that would reduce the estimation error $e(n)$. An example of the ML techniques is multiple linear regression method that can be used to predict the modeling parameters of the channel following the ABG model that was introduced by the 3GPP.
Multiple linear regression is a supervised learning and the goal is to infer and predict a function by reducing the error using the training data to predict the target by using machine learning method that perform a better estimation $\hat{f}$. In our scenario, using dataset that divided into two part, training and testing where the training set consists of 70 \% and the test is 30\% and other as mentioned in \cite{AMolisch2018}. The multiple linear equation can be seen as follow \cite{Freedman}:
\begin{equation}
f_i(x)=\beta_0+\beta_1 X_1+ ... +\beta_i X_i+\varepsilon
\end{equation}
The estimated of the response variable $Y$ (PL) using $X$ as CSI and minimizing the erroneous follow:
\begin{equation}
\hat{y_i}=\hat{\beta_0}+\hat{\beta_1}\hat{X_1}+ ... +\hat{\beta_i} X_i
\end{equation}
Then, to maximize the estimation method, the coefficient of equation 16 has to be minimized to obtain a lower difference between the real and the estimation equation. The residual error of the regression estimation can be obtained using the below equations \cite{Saud}.
\begin{equation}
Minimize\sum_{i=1}^{n}e^2_i 
\end{equation}
\begin{equation}
e_i^2=\arg min \sum_{i=1}^{n} (y_i - \hat{y_i})^2
\end{equation}
\begin{equation}
(\beta_0, ... ,\beta_i)=\arg min_{\beta_0 ... \beta_1} [\frac{1}{n} \sum_{i=1}^{n} (y_i -\hat{\beta_0}-\hat{\beta_1}X_1- ... -\hat{\beta_i}X_i)^2
]
\end{equation}
Then, estimating the of the slope and variance can be shown below \cite{KC2019}.
\begin{equation}
\hat{\beta_1}=(X^TX)^{-1}X^Ty
\end{equation}
Using the above estimate, we can infer the variance $\sigma^2$ in order to find the shadow fading parameter of the Close-in and ABG models ()equations 7, 10 and 11).
\begin{equation}
\hat{\sigma^2}=\frac{1}{L-1}(y-X\hat{\alpha})^{T}(y-X\hat{\alpha})
\end{equation}
Obtaining these parameters, lead us to build the path loss model that is comparable to model in earlier chapter. Moreover, other machine learning algorithms can be used to develop an alternative procedure to enhance the estimation and solve other channel modeling issues. 
\section{Data}
Investigation the channel modeling would allow other applications to interchange data to make the communication between them more precisely. Getting machine learning involved in a measurement from simulation or campaign with ML algorithms will provide homogeneous works and unbiased results. Machine learning algorithms investigate the feature of wireless channels deeply in MmWave. Machine learning is used to improve the performance and reduce complexity. By using a measurements dataset, ML methods would be applied to enhance the accuracy or to interpret/extend non-measured scenarios. Channel modeling parameters are generated by measurements campaigns or simulations and the propagated signal through a channel that gets disturbed by fading which leads to MPCs.The highest MPCs is the strongest link and from there, channel parameters can be obtained to create a dataset. 
We modified an open source Matlab simulation that was provided by New York University throughout their wireless lab \cite{NYUSIM} \cite{NYUSIM2} to meet our specifications. Then, that simulation was used to obtain a sufficient amount of data to enhance the accuracy of the models. Then, we purposed methods of using multiple of machine learning techniques and the generated data and then do the interpretation of performance comparison between the algorithms to check the path loss model as shown in the results section. The regression techniques in consideration include linear regression methods. Python was used to preform   the data analysis.
\begin{table}[!h]
	\caption{Channel Measurement Parameters.}\label{tab1}
	\begin{tabular}{| m{5 cm} | m{2cm}|} 	
		\hline
		\bfseries  Parameters &   \bfseries Values \\
		\hline
		Distance (m)& 1-40 \\
		\hline
		Frequency (GHz) & 28 \\
		\hline
		Bandwidth (MHz) & 800\\
		\hline
		TXPower (dBm) & 300\\
		\hline
		Scenario & UMi \\
		\hline
		Polarization & Co-Pol \\
		\hline
		TxArrayType & ULA \\
		\hline
		RxArrayType & ULA \\
		\hline
		Antena & SISO \\
		\hline
		Tx/Rx antenna Azimuth and Elevation (red) & 10\\
		\hline
	\end{tabular}
\end{table}
Table \rom{1}., exhibits that the channel measurement parameters of the data raw that was used for this paper.
Regression is considered the main methods to investigated the relationship between the channel features \cite{KC2017}. With the glory of having a large amount of data, the behaviour of the wireless channel modeling becoming more interesting and obvious to obviate the complexity. The step following cleaning the data is applying the machine learning scheme to stars the learning processes. Then, a model can be used for predicting the path loss and evaluating the model will be accomplished by Mean Absolute Error, Mean Squared Error, Root Mean Squared Error and R-square as shown in the result section.
\section{Results}
Checking how significant the data is perform by using residual plots, where plot as shown in figure 2 can demonstrate how the data distributed among the horizontal line and it appears reasonably random. Thus, it confirms the data used for regression is unbiased. Residuals method is used to forecast errors which can be obtained by subtracting the forecast from the expected values.
\begin{figure}[!h]
	\centering
	\setlength{\abovecaptionskip}{0.1cm}
	\setlength{\belowcaptionskip}{-0.3cm}
	\includegraphics[width=0.45\textwidth]{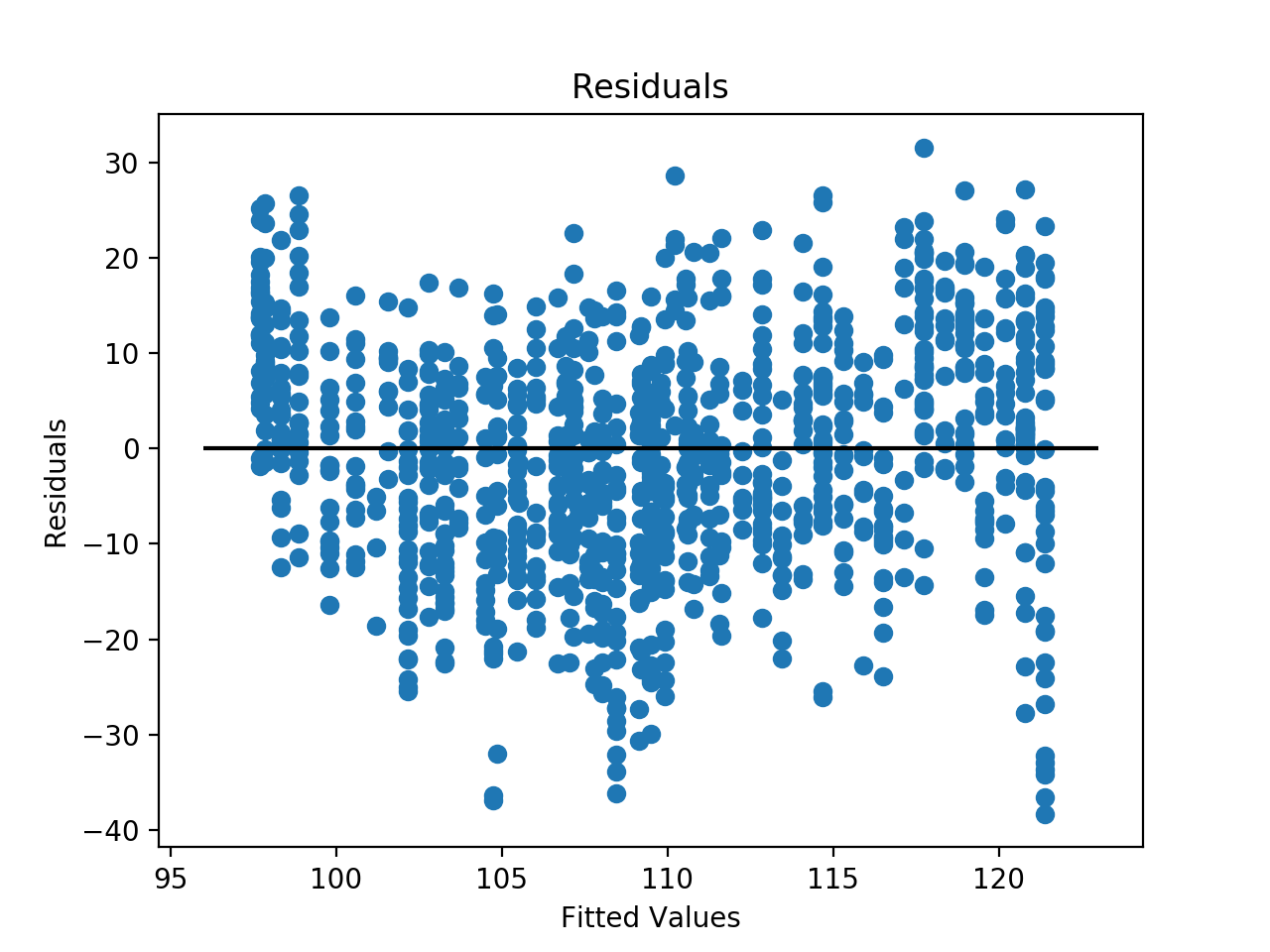}
	\renewcommand\figurename{Fig}
	\caption{Residuals Plot.}
	\label{Fig5}
\end{figure}
Figure 2 shows the prediction of path loss using the linear regression method and evaluating this model is shown in table \rom{3}. While table \rom{2} explores the coefficient parameters of the Linear Regression (LR), Multiple Linear Regression (MLR1) and Multiple Linear Regression (MLR2). The second model (MLR1) implemented with only three wireless channel feature while the third model used eight features.
\begin{figure}[!h]
	\centering
	\setlength{\abovecaptionskip}{0.1cm}
	\setlength{\belowcaptionskip}{-0.3cm}
	\includegraphics[width=0.45\textwidth]{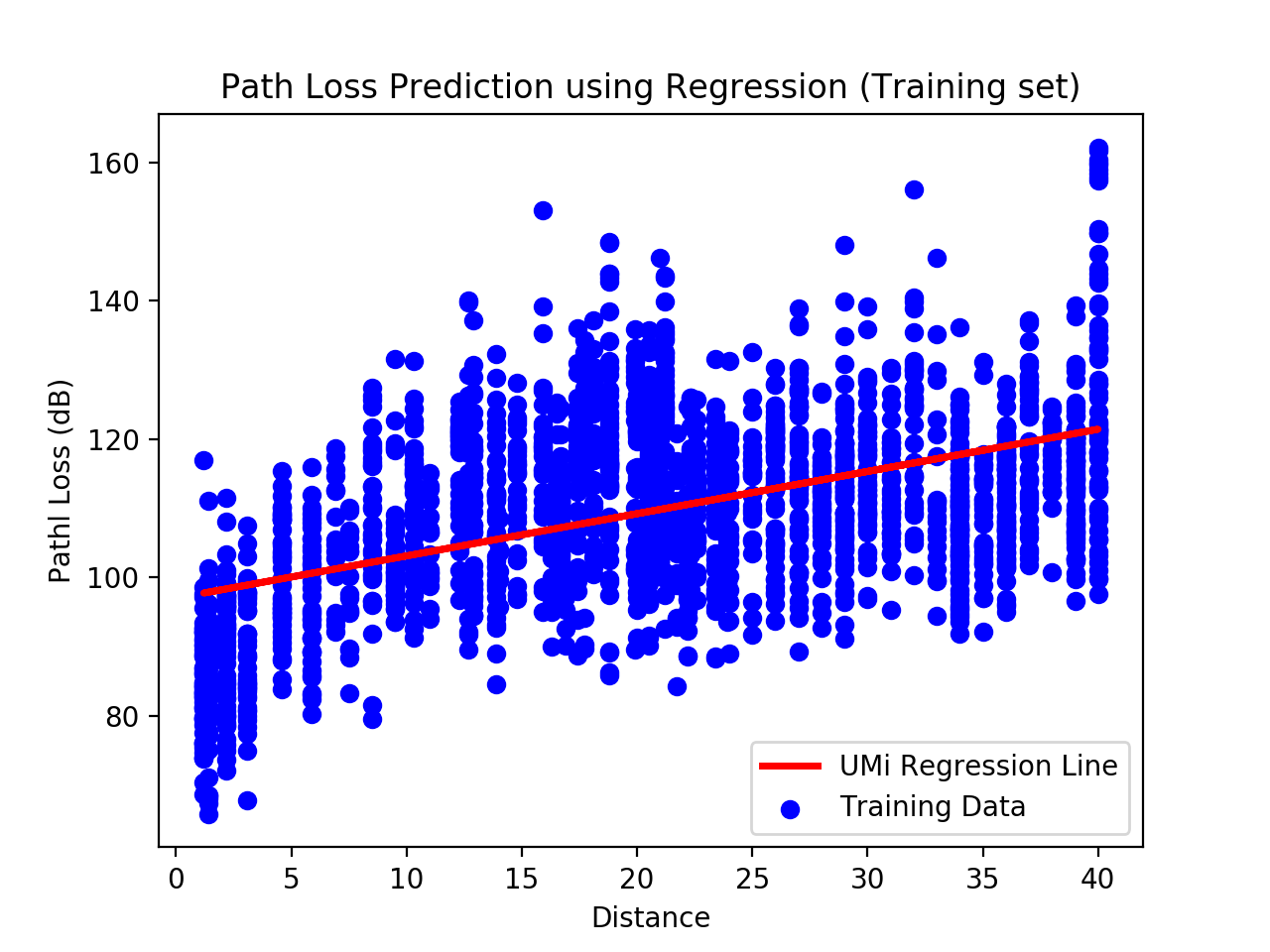}
	\renewcommand\figurename{Fig}
	\caption{Path Loss Prediction Using linear Regression algorithm.}
	\label{Fig5}
\end{figure}
Figure 3 demonstrates two model urban micro (Uma) and urban micro (Umi) of wireless communication. The date belongs to umi that was used to generate a regression line, while the regression line is the uma model that was obtained from uma scenario and then applied to the umi.
\begin{figure}[!h]
	\centering
	\setlength{\abovecaptionskip}{0.1cm}
	\setlength{\belowcaptionskip}{-0.3cm}
	\includegraphics[width=0.45\textwidth]{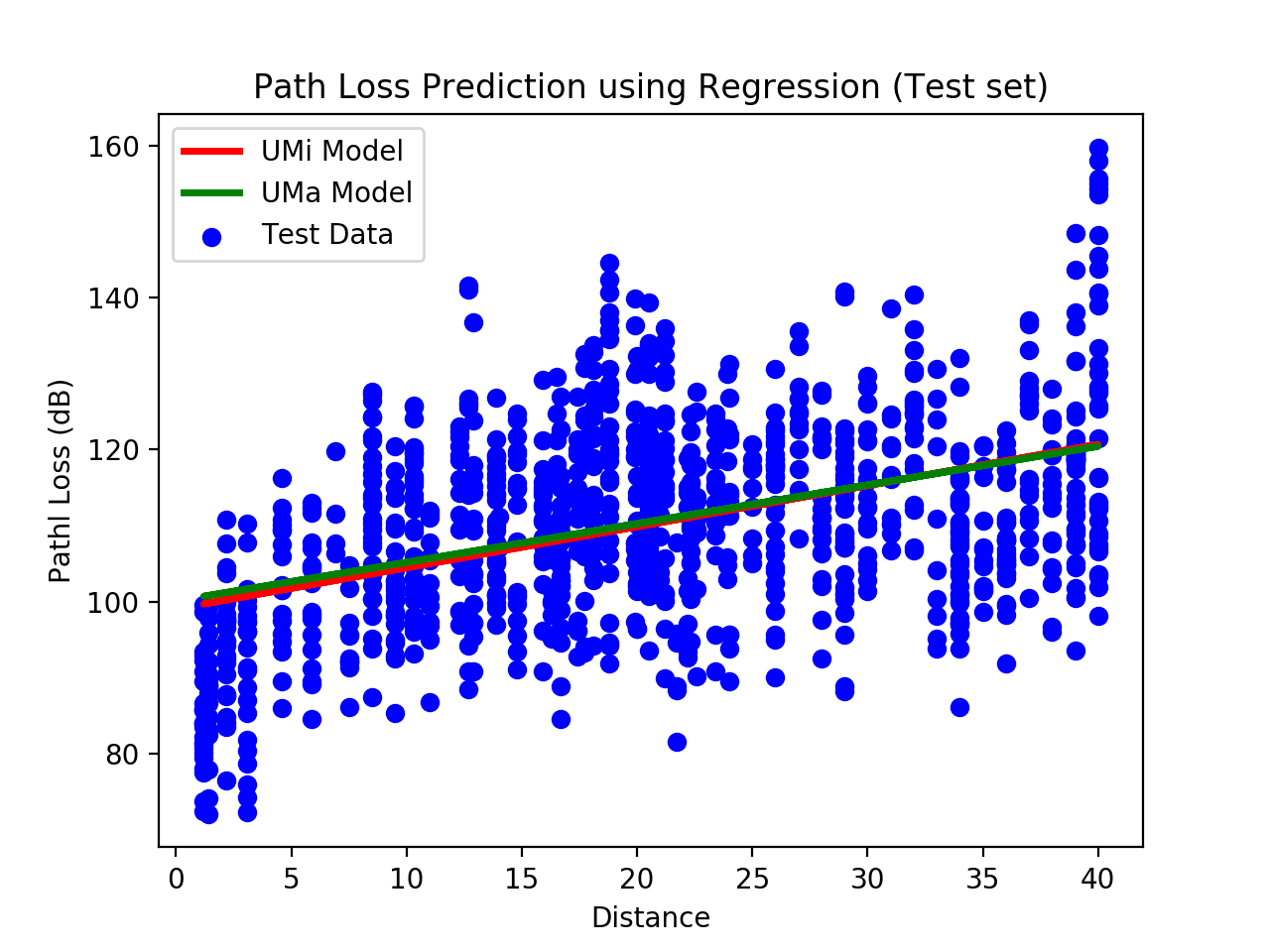}
	\renewcommand\figurename{Fig}
	\caption{Path Loss Prediction Using linear Regression algorithm.}
	\label{Fig5}
\end{figure}
 With the usage of different wireless channel features, table \rom 2 illustrate an adequate work by comparing the results of applying the model that were obtained from micro/urban environments with table \rom 1 specifications and applied it in the macro/urban communication. Thus, the wireless channel measurements can be reduce by applying a model from a single environment to others by applying machine learning techniques that can learn the logic.
\begin{table}[!h]
	\caption{Communication Scenarios Comparison .}\label{tab1}
	\begin{tabular}{| m{3.5 cm} | m{2.10cm}|  m{1.60cm}|} 	
		\hline
		\vspace{0.2mm}
		\bfseries Environment Scenario &  \bfseries  UMi &  \bfseries UMa  \\
		\hline
		\textbf{MAE} &  8.92& 6.66 \\
		\hline
		\textbf{MSE} & 126.60 &	 74.32   \\
		\hline
		\textbf{RMSE}& 	11.25 &	8.62  \\
		\hline
		\textbf{R Square}& 	0.21 & 0.533 \\
			\hline
		\textbf{Confidence}& 	0.21 & 0.533 \\
		\hline
	\end{tabular}
\end{table}
\begin{table}[!h]
	\caption{Channel Measurement Parameters for UMi Communication.}\label{tab1}
	\begin{tabular}{| m{3.55 cm} | m{1.150cm}|  m{1.150cm}|  m{1.150cm}|} 	
		\hline
		\vspace{0.2mm}
		\bfseries Test &  \bfseries  LR &  \bfseries MLR &\bfseries MLR \\ 
		\hline 
		\textbf{T-R Separation Distance (m)	} & 0.56& 0.46	& 0.48\\
		\hline
		\textbf{Time Delay (ns)} & - &	 -0.08  &-0.09 \\
		\hline
		\textbf{Received Power (dBm)}& - &	-0.69  &-0.69\\
		\hline
		\textbf{RMS Delay Spread (ns)}& 	-& 	-& 	0.29\\
		\hline
		\textbf{Elevation AoD (degree)} &  	-& 	-&-0.10 \\
		\hline
		\textbf{Azimuth AoD (degree)} & 	-&		-  &-0.002\\
		\hline
		\textbf{Azimuth AoA (degree)}& 		- &		-   &-0.004\\
		\hline
		\textbf{Elevation AoA (degree)}& 		-& 	- &	-0.001\\
		\hline
	\end{tabular}
\end{table}
\begin{table}[!h]
	\caption{Linear Regression Model.}\label{tab1}
	\begin{tabular}{| m{3.55 cm} | m{1.150cm}|  m{1.150cm}|  m{1.150cm}|} 	
		\hline
		\vspace{0.2mm}
		\bfseries Environment &  \bfseries  $\alpha$ &  \bfseries $L_0[dB]$ &\bfseries $X_\sigma{}[dB]$ \\ 
		\hline 
		\textbf{Outdoor Micro Urban	} &  9.7 & .61	& 13.6 \\
		\hline
	\end{tabular}
\end{table}
As can be followed by equation 4 and 5, the estimated path loss model shown as for linear regression and a single feature loss $L_0[dB]$ as the separated distance.
\begin{equation}
\hat{PL}= \alpha + 10 log L_0[dB](d) + X_\sigma{}[dB]
\end{equation}
While for the model for multiple regression that consists of multiple wireless channel features loss $L_N[dB]$ as equation 7 proved shown as:
\begin{equation}
\hat{PL}= \alpha + L_0[dB]+ L_1[dB]+ ... +L_9[dB]+ X_\sigma{}[dB]
\end{equation}
Both parameters of above two equations can be obtained from tables \rom{3} and \rom{4}. Moreover, using the statistical parameters Mean Absolute Error (MAE), Mean Squared Error (MSE), Root Mean Squared Error (RMSE) and R square ($R^2$) values level to achieve the significant of the predicted or the used model. RMSE is the square root of MSE and used to check the accuracy of the wireless channel propagation predication where it measures the differences between the predicted and observed model where zero value indicates the fit is optimum \cite{Ai2018}. Furthermore, these parameters can be used to validate the significance and check the accuracy of the proposed models.
Table \rom{3} illustrates the analysis of this journey, where there are three model that can predict the path loss of an outdoor Micro environment using a 28 GHz. The features that are used in the second model are T-R Separation Distance (m), Time Delay (ns) and 'Received Power (dBm). While the third model's features are T-R Separation Distance (m), Time Delay (ns) and 'Received Power (dBm), RMS Delay Spread (ns), Elevation AoD (degree), Azimuth AoD (degree), Azimuth AoA (degree) and Elevation AoA (degree). Then these models are evaluated using Mean Absolute Error, Mean Squared Error, Root Mean Squared Error and R-square. From table \rom{3}, multiple linear regression with eight features performs the best among other models which leads to increasing the feature enhance the prediction but until the model reaches to the steady state. R square particularly presents how the models improved with increasing the number of channel variable which provide an acceptable prediction result.
\begin{table}[!h]
	\caption{Micro Urban Channel Measurement Parameters.}\label{tab1}
	\begin{tabular}{| m{1.125 cm} | m{2.210cm}|  m{1.60cm}|  m{1.60cm}|} 	
		\hline
		\bfseries Test &  \bfseries  Linear Regression &  \bfseries Multiple Linear Regression (3 Feature) &\bfseries Multiple Linear Regression (7 Feature)\\
		\hline
		\vspace{0.2mm}
		\textbf{MAE} &  8.92& 6.66	&5.10 \\
		\hline
		\vspace{0.2mm}
		\textbf{MSE} & 126.60 &	 74.32   &44.51 \\
		\hline
		\vspace{0.2mm}
		\textbf{RMSE}& 	11.25 &	8.62   &6.67\\
		\hline
		\vspace{0.2mm}
		\textbf{R Square}& 	0.21 & 0.533 &	0.72\\
		\hline
	\end{tabular}
\end{table}
\section{Conclusion}
Inaccuracy, complexity and number of measurement of the wireless communications have been not solvable issues for past decay. This paper presented a new ML procedure to overcome these issue with the assist of machine learning techniques. The traditional solutions base on top wireless communication organizations have not sufficiently overcame these issue and with the new era of big data, it's the time to resolve them base on machine learning algorithms.  A new approach of applying supervise learning to model the wireless channel. We have used regression techniques to defeat the channel modeling issues. Using the data of certain communication environment, we were able to predict the model of the new communication scenario. Thus, the required number of measurements and the complexity have been reduced.
\section{Acknowledgement}
Saud Aldossari expresses a great appreciation to Prince Sattam bin Abdulaziz University for their support of providing a scholarship. 

\balance 

\bibliographystyle{IEEEtran}
\end{document}